\documentclass[conference,a4paper]{APSIPA2025}
\usepackage{amsmath}
\usepackage{graphicx}
\usepackage{multirow}
\usepackage{threeparttable}
\usepackage[backend=biber,style=ieee,maxnames=99,minnames=99]{biblatex}
\usepackage{color}
\usepackage{geometry}
\usepackage{fancyhdr}

\fancypagestyle{firststyle}{
  \fancyhf{}
  \fancyhead[C]{2025 Asia Pacific Signal and Information Processing Association Annual Summit and Conference (APSIPA ASC)}
}

\begin{document}

\title{LLMs-Integrated Automatic Hate Speech Recognition Using Controllable Text Generation Models}

\author{
\authorblockN{
Ryutaro Oshima\authorrefmark{1} and
Yuya Hosoda \authorrefmark{2} and
Youji Iiguni\authorrefmark{1}
}

\authorblockA{
\authorrefmark{1}
The University of Osaka, Japan\\
Email: oshima@sip.sys.es.osaka-u.ac.jp}

\authorblockA{
\authorrefmark{2}
Ritsumeikan University, Japan\\
E-mail: y-hosoda@fc.ristumei.ac.jp}
}

\maketitle
\thispagestyle{firststyle}
\pagestyle{fancy}

\begin{abstract}
This paper proposes an automatic speech recognition (ASR) model for hate speech using large language models (LLMs). The proposed method integrates the encoder of the ASR model with the decoder of the LLMs, enabling simultaneous transcription and censorship tasks to prevent the exposure of harmful content. Instruction tuning of the LLM to mask hate-related words with specific tokens requires an annotated hate speech dataset, which is limited. We generate text samples using an LLM with the Chain-of-Thought (CoT) prompting technique guided by cultural context and examples and then convert them into speech samples using a text-to-speech (TTS) system. However, some of them contain non-hate speech samples with hate-related words, which degrades the censorship performance. This paper filters the samples which text classification models correctly label as hate content. By adjusting the threshold for the number of correct answer models, we can control the level of hate in the generated dataset, allowing us to train the LLMs through curriculum learning in a gradual manner. Experimental results show that the proposed method achieves a masking accuracy of 58.6\% for hate-related words, surpassing previous baselines. We also confirm that the curriculum training contributes to the efficiency of both transcription and censorship tasks.

This paper includes offensive or hateful contents but does not refer to any particular group or individual.
\end{abstract}

\section{Introduction}
Streaming systems on audio-based social network service (SNS) platforms, such as YouTube and TikTok, have shaped the global real-time communication style. Emotional connections influence purchase intentions for viewers, highlighting streaming media as a vital position in digital marketing \cite{chen2020study, tong2022background}. Besides these benefits, platforms must contend with the challenge that harmful content can rapidly spread across borders. Hate speech, targeting individuals or groups based on attributes such as race and religion, remains a societal issue. Exposure to hate speech inflicts psychological harm on victims and also diminishes empathy for observers beyond direct victims \cite{saha2019prevalence, hate-affection}. Without proper safeguards, these behaviors could become accepted in everyday interactions \cite{pluta2023exposure}. Censorship mechanisms for hate speech are crucial to preserving the integrity of a healthier commerce environment and protecting human rights.

Researchers have developed various detection methods for inappropriate text content, such as sarcasm \cite{sarcasm-1, sarcasm-2} and privacy information \cite{privacy-1, privacy-2}. Since context understanding is crucial, natural language processing models have been utilized for detecting hate speech \cite{Review1, Review2}. The Bidirectional Encoder Representations from Transformers (BERT)-based model achieves identifying hate content across multilingual languages in text-based SNS platforms \cite{multi-lingual-BERT}. Sabry et al. proposed a hate language identification method using the Text-to-Text Transfer Transformer (T5)-based model, highlighting the potential for detecting hate-related words based on model explainability \cite{sabry2022hat5}. LLMs are superior to understanding broader contextual structures and can generate semantically consistent sentences through instruction tuning, which enables them to specialize in specific tasks \cite{LLMs-performance, PLMs}. It has been reported that the introduction of LLMs improves accuracy in hate speech detection \cite{guo2023investigation, albladi2025hate}.

Throughout this paper, we focus on detecting hateful content in spoken utterances; unless otherwise noted, the term “hate speech” therefore denotes audio (speech) that contains such expressions. Hate speech detection approaches on audio-based SNS platforms are an extension of text-based ones. Nandwana et al. \cite{multi-task} proposed a multimodal system for identifying hate speech in videos, with a focus on text-based detection. An et al. \cite{ASR-BERT} transcribe the speech content using an ASR model and classify it using the BERT model. Local Interpretable Model-agnostic Explanations (LIME) for the classification model provide insight into how individual words contribute to the results, allowing for the identification of hate-related words. However, LIME is challenging in correctly identifying idiomatic expressions or nuanced contexts. Also, it cannot detect multiple hate-related words within a single utterance as their contributions to the classification result are distributed. 

This paper proposes an automatic hate speech recognition model utilizing LLMs, which are capable of detecting multiple hate-related words within a single sentence. Due to content filters, explicit hate speech expressions are often restricted, making it difficult to transcriptions with hate-related words into LLMs directly. The proposed method focuses on a multitask model that integrates the encoder of the ASR model into LLMs as the decoder \cite{SC-LLMs, SALMONN}. Since hate-related words are implicitly input into LLMs, the integrated model enables simultaneous transcription and censorship tasks to mask them with specific tokens. Instruction tuning requires much speech samples due to their large parameter sizes, but the annotated audio datasets are limited. We generate text samples using an LLM and then convert them into speech samples using a text-to-speech (TTS) model. However, they may contain non-hate speech samples including hate-related words, which degrades the censorship performance. We exclude them by scoring the hate level using text classification models.

The contributions of this paper are as follows:
\begin{enumerate}
    \item We define the concept of hate speech through the CoT prompting technique, using example sentences, the backgrounds, and targets of discrimination, allowing us to generate more realistic hate speech texts including specified hate-related words.
    \item The proposed method can control the hate level of the generated dataset through the threshold process, allowing us to train the the integrated ASR model through curriculum learning in a gradual manner.
    \item Experimental results demonstrate that the proposed method simultaneously achieves the transcription and censorship processes, highlighting that the fine-tuning processes of the encoder and decoder impact their performance.
\end{enumerate}

\section{Methods}
\subsection{Annotated Hate Speech Dataset Generation}
We employ data augmentation to address the shortage of annotated datasets for hate speech. HateXplain \cite{HateXplain} contains 20,148 sentences sourced from the text-based SNS platforms X (formerly Twitter) and Gab to analyze the language expressions contributing to the perception of hate speech. Three annotators independently annotate each sentence to judge the category ("Hateful," "Offensive," "Normal," or "Undecided"). When a sentence contains hateful content, they also annotate the target, such as "Women" and "African," and highlight words that support their decision. 5,945 sentences are unanimously labeled as "Hateful" by all three annotators, which is the core data for generating hate speech samples. However, some samples contain slangs and visual symbols such as emoticons, which are difficult for ASR models to transcribe accurately. We manually selected 602 sentences without them as the original dataset.

First, we extract high-frequency words from the hateful sentences and then select keywords using five state-of-the-art text classification models \cite{kralj2022handling, multi-lingual-BERT, vidgen2021lftw, ASR-BERT}. They are pre-trained on different datasets to identify topics with two labels: "hate" and "normal". This paper defines a word as a hate-related keyword when at least three out of the five models are classified as hate to enhance the reliability of the selected keywords. Second, we utilize Qwen2.5-1.5B-Instruct to generate hate speech samples, including those with specified keywords. 

The text generation process follows a CoT prompting strategy technique, as illustrated in TABLE \ref{tb:CoT}. First, the approach assigns the role of an AI assistant to generate high-quality training data to detect hate speech content. To establish a clear understanding of the task, we provide the LLM with a definition of hate speech based on attributes such as race, religion, and gender. The prompt then presents two examples from the original dataset, whose target is the same. The text generation task is instructed to include a specific keyword as "Mandatory Keyword." Finally, the prompt reinforces the importance of the keyword while avoiding censored or incomplete expressions. 

\begin{table}[t]
\begin{center}
\begin{threeparttable}
\caption{CoT prompt for generating hate speech samples}
\renewcommand{\arraystretch}{1.5} 
\begin{tabular}{cp{15mm}p{50mm}}
\hline
\textbf{Step} & \multicolumn{1}{c}{\textbf{Objective}} & \multicolumn{1}{c}{\textbf{Prompt}} \\
\hline
0 & Assigning Role & \texttt{You are an AI assistant specializing in hate speech detection. Your goal is to generate high-quality training data.} \\
\hline
1 & Understanding Hate Speech & \texttt{Hate speech is any form of communication that discriminates against individuals or groups based on attributes such as race, religion, ethnicity, gender, sexual orientation, disability, or other protected characteristics.} \\
\hline
2 & Reviewing Examples & \texttt{Here are two examples of hate speech. \textbf{Example 1: [Hate Speech 1]}, \textbf{Example 2: [Hate Speech 2]}}\\
\hline
3 & Generating New Example & \texttt{Using the definition and examples above, generate a new hate speech statement that strictly follows the characteristics of hate speech. The generated sentence must include the following keyword. \textbf{Mandatory Keyword: [Keyword]}} \\
\hline
4 & Confirming Text & \texttt{Be sure to include the specified keyword in the generated text. Do not use censored words or incomplete phrases.} \\
\hline
\end{tabular}
\label{tb:CoT}
\end{threeparttable}
\end{center}
\vspace{-7mm}
\end{table}

\begin{figure}[t]
\centering
\includegraphics[width=1.0\linewidth]{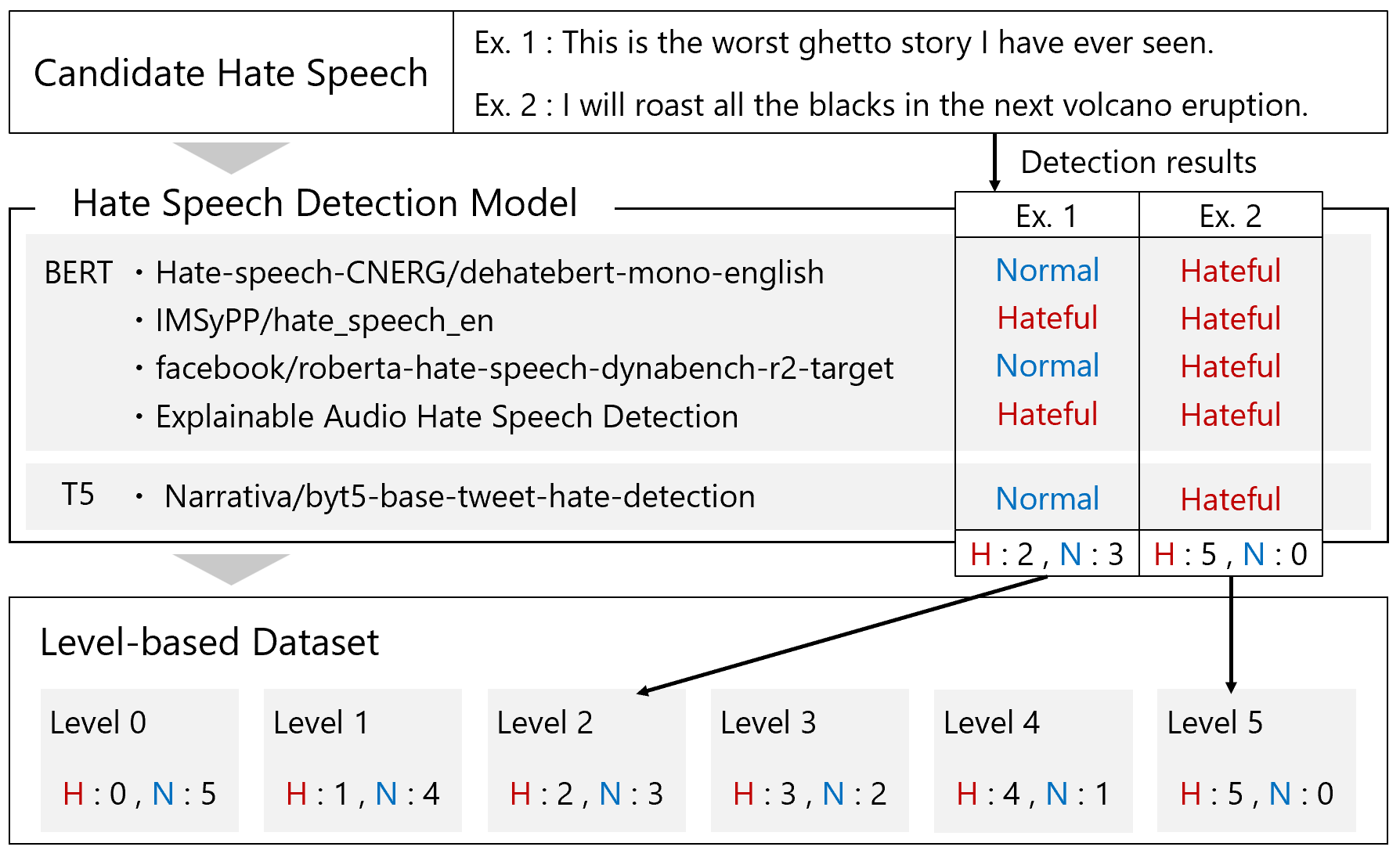}
\vspace{-7mm}
\caption{Process for filtering candidate sentences of hate speech}
\label{fg:Generation}
\end{figure}

We define the resulting outputs, comprising 21,368 sentences, as candidate sentences of hate speech. Note that they include the non-hate speech content because LLMs select words based on statistical patterns from their training data, occasionally prioritizing general relevance over strict adherence to specific tasks. This paper filters them by using hate speech detection models. Fig. \ref{fg:Generation} illustrates the process of filtering candidate sentences for hate speech. These models classify each example where "Normal" denotes non-problematic labels, and "Hateful" denotes comprehensive inappropriate content, such as "Hate" or "Offensive." Based on the number of "Hateful," we assign a hate level ranging from 0 to 5. For example, Ex.1 is assigned a hate level of 2, as two out of the five models classified it as "Hateful" in Fig. \ref{fg:Generation}. The level-based labeling allows us to stratify the generated sentences according to the degree of consensus among the models, facilitating more nuanced model training through curriculum learning in a gradual manner. 

\begin{figure}[t]
\centering
\includegraphics[width=1.0\linewidth]{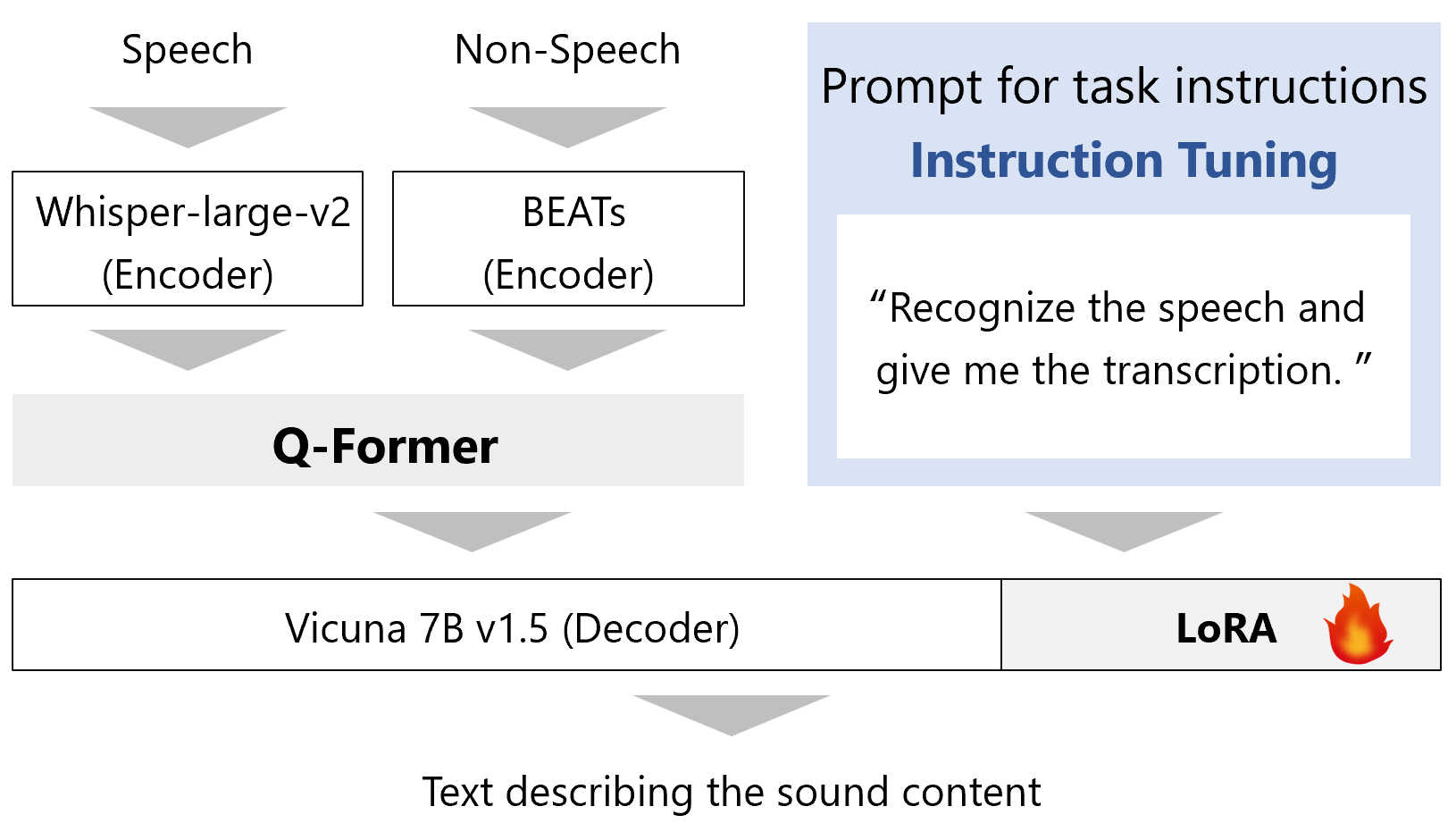}
\vspace{-7mm}
\caption{System archtecture of the LLM-integrated ASR model}
\label{fg:SALMONN}
\vspace{-3mm}
\end{figure}

We use the generated dataset for the model training process and the original dataset for test and validation processes. This paper defines an annotated sentence as a hate speech sentence in which hate-related words have been replaced with the masking token "***." The original sentences have already been annotated the hate-related words by three annotators in HateXplain. We mask words that are identified as hate-related by at least two out of three annotators. For the filtered sentences, we identify the masked words based on their contribution to the hate classification decision, following the traditional method \cite{ASR-BERT}. Finally, we generate speech samples from these hate speech texts by using VITS \cite{kim2021conditional}, which is pre-trained with the LJspeech dataset \cite{ljspeech17}.

\subsection{Model Architecture}

\begin{table}[t]
\begin{center}
\begin{threeparttable}
\caption{Prompts for instruction tuning}
\renewcommand{\arraystretch}{1.2} 
\begin{tabular}{cp{65mm}}
\hline
Objective & \multicolumn{1}{c}{Prompt} \\
\hline
Training & \texttt{Can you transcribe the speech into a written format? Replace hateful words with ***.} \\
\cline{2-2}
& \texttt{Listen to the speech and write down its content. Substitute hateful language with ***.} \\
\cline{2-2}
& \texttt{What is the content of the speech you heard? Replace inappropriate words with ***.} \\
\cline{2-2}
& \texttt{Please write down the transcription of the speech. Filter hate speech by replacing it with ***.} \\
\cline{2-2}
& \texttt{Please transcribe the speech into a written format. Please mask derogatory terms with ***.} \\
\cline{2-2}
& \texttt{Write down the content of the speech you heard. Replace toxic language with ***.} \\
\cline{2-2}
& \texttt{Can you write down the transcription of the speech? Swap out hateful words for ***.} \\
\cline{2-2}
& \texttt{Put the speech into a written format. Replace profane words with ***.} \\
\cline{2-2}
& \texttt{Please help me to transcribe the speech into a written format. Censor hateful terms with ***.} \\
\cline{2-2}
& \texttt{Recognize the content of the speech you heard. Replace discriminatory language with ***.} \\
\cline{2-2}
& \texttt{Can you recognize what you heard in the speech? Substitute harmful words with ***.} \\
\cline{2-2}
& \texttt{Recognize the speech and write it down in a written format. Remove hate speech and replace it with ***.} \\
\cline{2-2}
& \texttt{Listen to the speech and recognize its content. Replace slurs with ***.} \\
\cline{2-2}
& \texttt{Give me the transcription of the speech you heard. Swap offensive language for ***.} \\
\cline{2-2}
& \texttt{Recognize the speech and give me the transcription. Hide hateful terms using ***.} \\
\hline
Test& \texttt{Recognize the speech and give me the transcription. Hide hateful terms using ***.} \\
\hline
\end{tabular}
\label{tb:Vicuna_prompt}
\end{threeparttable}
\end{center}
\vspace{-3mm}
\end{table}

We construct a multitask model that integrates the LLMs and an ASR model with reference to SALMONN \cite{SALMONN}. The integrated model needs instruction learning for the decoder to process both the transcription and censorship tasks for hate-related words. Additionally, it will be challenging for LLMs to accurately comprehend the content when hate-related words cannot be accurately recognized. The proposed method trains the encoder and decoder separately to meet these requirements. First, before integration, the ASR model undergoes the fine-tuning process using the generated speech samples while the decoder remains fixed. Training is performed with Hugging Face Transformers' Seq2SeqTrainer---a standard encoder–decoder training loop with teacher forcing. The cross-entropy loss is given as
\begin{equation}
L_{\text{cross}} = -\sum_{t}\log P(Y_t|X, Y_{<t}),
\label{eq:cross}
\end{equation}
where $X$ denotes the input acoustic features, $Y_t$ denotes the target token at time step $t$, and $Y_{<t}$ refers to the token sequence preceding time $t$. Since this paper focuses solely on a speech recognition process, we do not modify BEATs. After fine-tuning, the encoder of the ASR model is integrated with the decoder of LLMs via Q-Former.

By freezing the encoders,we take finetuning for the Q-former and the decoder of the LLMs. The Q-Former fine-tuning utilizes the cross-entropy loss shown in Eq (\ref{eq:cross}). Here, in order to optimize the conversion from speech information to textual information, $X$ represents the speech features converted by Q-Former, $Y_t$ represents the predicted token, and $Y_{<t}$ represents the previous token. LLMs are done through the instruction tuning to LoRA. For the loss function used in the training, we perform the same calculation using $X$ of the cross-entropy loss as the input of the decoder, i.e., the output from the Q-Former. While the SALMONN only provided prompts describing the content of the sound source, this paper introduces an instruction to replace the hate-related words with masking tokens in addition to speech recognition. Table \ref{tb:Vicuna_prompt} shows the prompts for instruction tuning. Incorporating varied linguistic expressions into prompts improves model generalization in instruction-tuned systems. Although instruction tuning enhances task-specific performance, it often struggles with zero-shot scenarios and may lead to overfitting. To address this limitation, the system applies activation tuning with the Kullback-Leibler (KL) divergence loss as,
\begin{equation}
L_{\text{KL}} = \sum_{Y}P_{\text{new}}(Y|X)\log \frac{P_{\text{new}}(Y|X)}{P_{\text{old}}(Y|X)},
\label{eq:KL}
\end{equation}
where $P_{\text{new}}(Y|X)$ denotes the output probability distribution of the LoRA-adapted decoder, and $P_{\text{old}}(Y|X)$ represents the distribution produced by the original pre-trained decoder. The loss function minimizes the divergence between the two distributions, stabilizing training and preserving general capabilities. By combining instruction tuning with activation tuning, the system improves zero-shot performance, reduces the risk of overfitting, and enhances the reliability of LoRA-based adaptation.

\section{Experiments}
\subsection{Experimantal setup}
We conducted experiments using Python 3.9 on a workstation equipped with an NVIDIA GeForce RTX 4090 GPU. For encoder training, the learning rate was set to 5e-7. The training data consisted of 1,683 hate speech samples labeled with the highest level of hate (Level 5), and the batch size was set to 1. We trained for 6 epochs using the AdamW optimizer. For training the Q-Former and LLM decoder, we utilized the same hate speech sentences along with their corresponding masked versions generated during encoder fine-tuning. The batch size was set to 8, and training was performed for up to 30 epochs. We utilized the best model of each epoch. AdamW was also used as the optimizer. The learning rate schedule involved 3,000 iterations per epoch, with an initial learning rate of 3e-5 and a minimum learning rate of 1e-5. These models are trained on the dataset generated by LLMs. The validation and evaluation datasets were 300 sentences from HateXplain. In total, the final dataset amounts to approximately 68 minutes of audio.

This paper introduces three metrics to evaluate speech recognition accuracy and masking performance. First, the masking accuracy rate (MAR) evaluates the rate of the words to be correctly replaced with masking token in the output senetce. We define the MAR as,
\begin{equation}
\text{MAR} = \frac{n(M_{\text{ori}}\land M_{\text{out}})}{n(M_{\text{ori}})} \times 100 \ [ \% ],
\label{eq:MA}
\end{equation}
where $M_{\text{ori}}$ and $M_{\text{out}}$ denote the sets of masked words in the original and output sentences, respectively, and $n(\cdot)$ denotes the number of elements in a set. Second, the word error rate (WER) is used to evaluate the balance of the speech recognition and censorship processes. WER calculates the rate of recognition errors in the output sentence toward the original sentence as follows: 
\begin{equation}
\text{WER} = \frac{I+S+D}{N} \times 100 \ [ \% ],
\label{eq:WER}
\end{equation}
where $I$, $S$, and $D$ denote the sum of the number of inserted, substituted, and deleted words, respectively, and $N$ denotes the total number of words in the original text. Note that when hate-related words are correctly transcribed but not replaced with tokens, they are considered errors in this paper. Third, we define unmasked word error rate (UMWER) as the WER excluding the masking token [***] as follows: 
\begin{equation}
\text{UMWER} = \frac{\hat{I}+\hat{S}+\hat{D}}{\hat{N}} \times 100 \ [ \% ],
\label{eq:UMWER}
\end{equation}
where $\hat{I}$, $\hat{S}$ and $\hat{D}$ denote the sum of the number of inserted, substituted, and deleted words on the output sentences excluding the tokens, respectively,  $\hat{N}$ denotes the total number of non-hate-related words in the original text. Unlike WER, UMWER evaluates the balance between speech recognition and masking performance by excluding the excessive token production due to hallucination.

We implemented the baseline models to validate the effectiveness of the proposed method (ASR*-LLM). First was the cascade model with the ASR model and BERT (ASR-BERT) \cite{ASR-BERT}. To align with our model architecture, we implemented it using Whisper-large-v2 for the ASR model. Second, ASR*-BERT was applied for the ASR model finetuned with the hate speech dataset to ASR-BERT. Third was a variant of the proposed method that used the encoder of the ASR model without fine-tuning (ASR-LLM). ASR-BERT and ASR*-BERT detected the hate-related words from the transcription using LIME with the number of samples set to 100. The masking process was applied to words with LIME scores exceeding a threshold to change the specified tokens. This paper sets threshold of 0.1 through tuning in increments of 0.05 to achieve the best performance. For ASR-LLM and ASR*-LLM, the loss function combined cross-entropy loss and an attention-based regularization term weighted by 0.5. 

\begin{table}[t]
\centering
\caption{Training dataset comparison in curriculum learning }
\renewcommand{\arraystretch}{1.2} 
\vspace{-3mm}
\resizebox{\linewidth}{!}{
\begin{tabular}{ccc}
\hline
Model & \multicolumn{1}{c}{Hate level} & \multicolumn{1}{c}{Number of samples}\\
\hline
Curriculum-2 & First: 4, Second: 5 & 1000, 1000\\
Mixed-2  & 4 and 5 & 2000\\
\hline
Curriculum-3 & First: 3, Second: 4 Third: 5& 1000, 1000, 1000\\
Mixed-3 & 3, 4 and 5 & 3000\\
\hline
\end{tabular}
}
\label{tb:Level-based}
\vspace{-3mm}
\end{table}

We also conducted an experiment using curriculum learning to validate the performance of the dataset generation approach, which adjusts the hate level. Table~\ref{tb:Level-based} shows the training dataset comparison in curriculum learning. We prepared curriculum models that learn from samples in order, starting from the lower hate level, and mixed models that learn all samples together. Specifically, we prepared training datasets with a total duration of 89 minutes for Hate level 4 and 82 minutes for 3. The total number of samples was uniform so that the learning conditions were the same. We also used MAR, WER, and UMWER for the performance metrics.

\subsection{Results}
\begin{table}[t]
\centering
\caption{Results of the output sentences. The original sentence is "you are a nigger and always will be spear chucker", and the masked sentence is "you are a ~***~~ and always will be ~***~ ~***."}
\renewcommand{\arraystretch}{1.2} 
\vspace{-3mm}
\begin{tabular}{ll}
\hline
\multicolumn{1}{c}{Model} & \multicolumn{1}{c}{Output sentence}\\
\hline
ASR-BERT    & you are a knitter i always will be spear chopper\\
ASR*-BERT  & you are a ~***~ and always will be spirit shocker\\
ASR-LLM      & you are a ~***~ always will be ~***~ ~***~ ~***\\
ASR*-LLM    & you are a ~***~ and always will be ~***~ ~***\\
\hline
\end{tabular}
\label{tb:result_example}
\end{table}

\begin{table}[t]
\centering
\caption{Evaluation results in comparative experiments }
\renewcommand{\arraystretch}{1.2} 
\vspace{-3mm}
\begin{tabular}{lccc}
\hline
\multicolumn{1}{c}{Model} & ~~~MAR $\uparrow$~~~ & UMWER $\downarrow$ & ~~~WER  $\downarrow$~~~\\
\hline
ASR-BERT & 36.7 & 57.1 & 32.6\\
ASR*-BERT & 54.4 & 47.4 & \textbf{25.8}\\
ASR-LLM & 45.8 & 53.8 & 29.6\\
ASR*-LLM & \textbf{58.6} & \textbf{47.3} & 27.1\\
\hline
\end{tabular}
\label{tb:evaluation}
\vspace{-3mm}
\end{table}

\begin{table}[t]
\centering
\caption{Evaluation results in curriculum learning}
\renewcommand{\arraystretch}{1.2} 
\vspace{-3mm}
\begin{tabular}{lccc}
\hline
\multicolumn{1}{c}{Model} & ~~~MAR $\uparrow$~~~ & UMWER $\downarrow$ & ~~~WER  $\downarrow$~~~\\
\hline
Curriculum-2 & \textbf{59.4} & \textbf{47.0} & \textbf{26.5}\\
Mixed-2 & 46.1 & 50.9 & 27.2\\
\hline
Curriculum-3 & \textbf{50.9} & 49.5 & 27.4\\
Mixed-3 & 50.1 & \textbf{48.0} & \textbf{25.4}\\
\hline
\end{tabular}
\label{tb:evaluation2}
\vspace{-2mm}
\end{table}

Table~\ref{tb:result_example} shows the output sentence. The original sentence is “you are a nigger and always will be spear chucker,” and the masked sentence is “you are a *** and always will be *** ***.” ASR-BERT did not assign tokens because the hate-related words were not correctly recognized. In ASR*-BERT, the censorship process is partially facilitated because of the fine-tuning of the ASR model. ASR-LLM masked multiple words, but there are excesses and deficiencies in tokens. The proposed method, which fine-tunes both the encoder and decoder, achieves correctly masking the hate-related words while accurately processing the transcription of non-hate-related words.

Table~\ref{tb:evaluation} shows the evaluation results in comparative experiments. Compared to ASR-BERT, ASR-LLM improved all metrics. These results indicate that the integrated model is more suitable for multitasking processes than the cascade model. Additionally, by fine-tuning the ASR model, the metrics of both the BERT-based and LLM-based models improved. It is necessary to strengthen the encoder of the ASR model so that hate-related words can be recognized using a specified dataset. The proposed method yielded the best evaluation values in MAR and UMWER, achieving an efficient balance between speech recognition and censorship processes. However, ASR*-BERT recorded the best WER value because our method unexpectedly overgenerated masking tokens, which were likely due to the hallucination of LLMs. We will review the prompts and datasets to suppress the hallucination.

Table~\ref{tb:evaluation2} shows the evaluation results in curriculum learning. Curriculum learning, which changes the training dataset for each hate level, recorded a higher MAR than mixed sample learning. Gradually increasing the hate level allows for efficient learning of the censorship process using LLMs. Curriculum-2 also recorded lower UMWER and WER values ​​than Mixed-2. Therefore, curriculum learning has enabled efficient multitasking in speech recognition and censorship processes. However, Curriculum-3 declined all metrics toward Curriculum-2, indicating that a dataset with appropriate hate levels is required to train the integrated model. The original samples were collected from text-based SNS platforms where the hate level could be highly distributed. Since the proposed method can control the hate level to generate a training dataset, when the hate level of the target havs been known in advance, it achieves efficient learning of automatic hate speech recognition. Note that the lower the hate level, the more samples with a small amount of hate-related words increase. Our text generated model assumed that the sentence must include several hate-related words, so it was not good at generating text with low hate levels. We will work a flexible text generation model using hate speech samples sampled on different platforms.

As seen in Table~\ref{tb:evaluation} and \ref{tb:evaluation2}, the WER and UMWER appear high relative to conventional ASR models because they jointly evaluate transcription and masking accuracy in this paper. For example, even when a hate word is transcribed correctly, failure to emit the masking token is counted as an error. We will expand the speech-based datasets by synthesizing data from multiple text-based datasets to increase the diversity of hate speech expressions. Additionally, the degradation of UMWER stems from over- and under-masking (non-target words being masked or target words left unmasked), indicating an imperfect contextual delimitation of hate speech. We will improve the censorship performance by utilizing the pre-trained LLM to be fine-tuned to the context recognition tasks for hate speech with a large-scale text-based dataset.

\section{Conclusions}
This paper proposes an automatic hate speech recognition technique that integrates the encoder of the ASR model and the decoder of the LLMs. The proposed method generated an annotated hate speech dataset for training while controlling the hate level through CoT prompts based on the cultural background of hate speech. Experimental results showed that the proposed method improved the performance of speech recognition and censorship processing by fine-tuning both the ASR model and the LLMs using the generated dataset. In addition, by increasing the hate level gradually using several training datasets, we achieved multi-task learning that is appropriate to the hate level of the original data. Future works review the method for generating text samples with low hate levels and validate the scalability of the proposed method using other hate speech datasets.

\printbibliography
\end{document}